\documentclass[showpacs,aps,superscriptaddress]{revtex4}
\usepackage{amssymb}
\usepackage{amsmath}
\usepackage{graphicx}
\usepackage{lscape}
\usepackage{booktabs}
\usepackage{epsfig}
\begin{document}
\title{ALP photo-production in the parton and hadron cascade model PACIAE}

\author{Dai-Mei Zhou}
\email[]{zhoudm@mail.ccnu.edu.cn}
\affiliation{Key Laboratory of Quark and Lepton Physics (MOE) and Institute of Particle Physics, Central China Normal University, Wuhan 430079,
            China}

\author{Zhi-Lei She}
\affiliation{School of Mathematics and Statistics, Wuhan Textile University, Wuhan 430200, China}

\author{Jian Cao}	
\affiliation{Key Laboratory of Quark and Lepton Physics (MOE) and Institute of Particle Physics, Central China Normal University, Wuhan 430079, China}

\author{An-Ke Lei}
\affiliation{School of Physics and Electronic Science, Guizhou Normal University, Guiyang 550025, China}

\author{Wen-Chao Zhang}
\affiliation{School of Physics and Information Technology, Shaanxi Normal University, Xi'an 710119, China}

\author{Hua Zheng}
\affiliation{School of Physics and Information Technology, Shaanxi Normal University, Xi'an 710119, China}

\author{Li-Lin Zhu}
\affiliation{College of Physics, Sichuan University, Chengdu 610064, China}

\author{Yu-Liang Yan}
\affiliation{China Institute of Atomic Energy, P. O. Box 275 (10), Beijing
102413, China}

\author{Ben-Hao Sa}
\email[]{sabhliuym35@qq.com, corresponding-author.}
\affiliation{China Institute of Atomic Energy, P. O. Box 275 (10), Beijing 102413, China}
\affiliation{Key Laboratory of Quark and Lepton Physics (MOE) and Institute of Particle Physics, Central China Normal University, Wuhan 430079, China}

\date{\today}

\begin{abstract}
In this work, the parton and hadron cascade model PACIAE is employed to
simulate the events of Au+Au collisions at $\sqrt{s_{NN}}=$200 GeV within the
60-80 \% centrality interval. Based on the abundant photons in final hadronic
state, the statistical-mechanics-based coalescence model DCPC is applied to
phenomenologically recombine post-freeze-out photon pairs into
axion-like particles (ALPs) for the first time. The coalescence probability is
constrained by the branching-ratio $B_r(a \rightarrow \gamma\gamma)=0.57$ under the
threshold of twice of $K$ meson mass. The ALP observables including yields,
transverse-momentum distributions and rapidity spectra are calculated for
several benchmark ALP-mass bins. These theoretical predictions are worthy to be investigated by the experiments.
\end{abstract}

\maketitle

\section{Introduction}

The Axion-like particle (ALP) is a pseudo-Nambu-Goldstone boson that appears
naturally in many extensions of the Standard Model. The strong Charge conjugation-Parity(CP) problem is
one of the most profound puzzles in particle physics. A solution to this
problem is to invent a new particle of (QCD) axion or, more generally, the ALP.
The (QCD) axion (with tiny mass) and the ALP (with mass treated as a free
parameter extending into MeV or even GeV range) are popular candidates of the
dark matter~\cite{hare,luzio,tamm}. The $e^+e^-$ collisions~\cite{jiang,abum}
and the nuclear-nuclear ultra-peripheral collisions (UPC)~\cite{knap} are
the most powerful experimental approaches to search for the production of ALP.

Another important issue, apart from the generation of ALP, is to study the
decay of ALP. For the ALP decay a few kinematically accessible two-body decay
channels below the threshold of twice of $K$ meson mass ($2m_K$)
are~\cite{tamm}:
\begin{enumerate}
\item If $m_a > 2m_{\pi}$ (=280 MeV):
\begin{itemize}
\item $a\rightarrow 2\pi$,
\item $a\rightarrow l^+l^-$,
\item $a\rightarrow \gamma\gamma$.	
\end{itemize}	
\item If $m_a < 2m_{\pi}$ but $m_a > 2m_e$ (=1 MeV):
\begin{itemize}
\item $a\rightarrow e^+e^-$,
\item $a\rightarrow \gamma\gamma$.	
\end{itemize}	
\item If $m_a < 2m_e$ \\ $a\rightarrow \gamma\gamma$.	
\end{enumerate} 		
Thus in antiparallel to the Belle II observation of the ALPs decaying to two
photons~\cite{abum}, one can use two photons with total energy less than the
$2m_e$ mass threshold to recombine an ALP by the statistical mechanics based coalescence model DCPC.

We employ the parton and hadron cascade model PACIAE 3.0~\cite{lei}
to simulate five million events for the 60-80 \% centrality Au+Au
collisions at $\sqrt{s_{NN}}=$200 GeV. Based on the photon list in each simulated event, two photons with total energy below the twice of electron mass ($2m_e$=0.001 GeV) threshold could be recombined an ALP
by the coalescence model DCPC. Unfortunately, such a low energy photon pair is
very rare. We are conscripted to change the tactics to recombine ALPs under
the twice of $K$ meson mass threshold with proper branching ratio of
$B_r(a\rightarrow \gamma\gamma)$. However, the value of this branching ratio
differs in the literature~\cite{jiang,abum,briv,baue,garc,tele}.  In this
work, we take a middle value of
$B_r(a\rightarrow \gamma\gamma)=$0.57~\cite{baue}.

As the ALP mass is treated as a free parameter~\cite{tamm} extending into the
MeV or even GeV range, we segment ALP mass into several groups and count their
yield, transverse momentum distribution, and rapidity spectrum, individually.
These theoretical predictions for the first time are worthy to be investigated by the experiments.

\section{PACIAE model and related DCPC model}

\begin{figure}[ht]
\centering
\epsfig{file=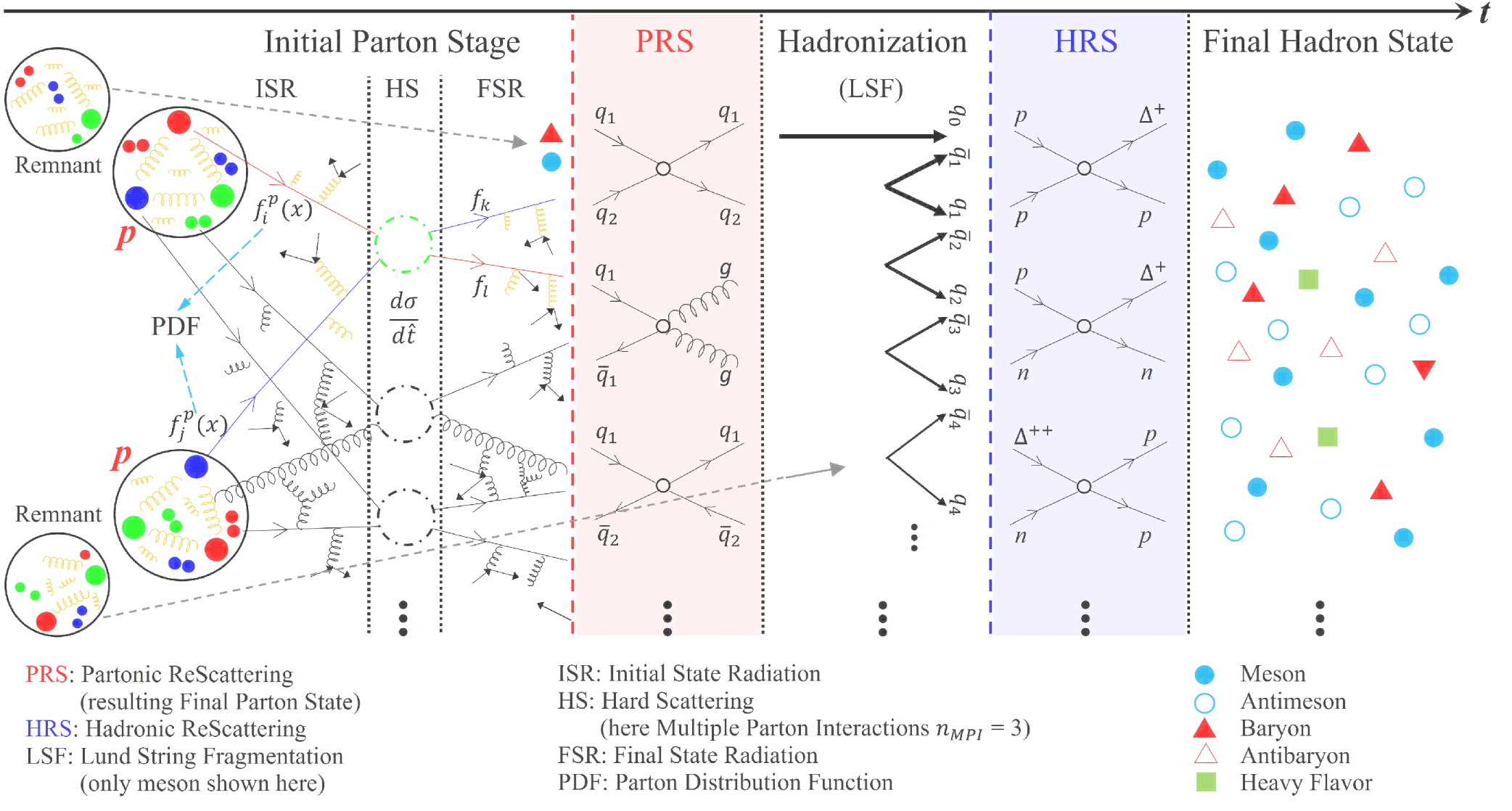,width=16.8cm,angle=0}
\caption{A sketch of the transport processes developed in a high energy pp collision.}
\label{phyrou0}
\end{figure}

For the high energy $pp$ collision the PACIAE model~\cite{lei} differs from
PYTHIA~\cite{soj1} in the implementations of the partonic and hadronic
rescatterings before and after hadronization, respectively.
Figure~\ref{phyrou0} is a sketch of the transport processes developed in
a $pp$ collision: A $pp$ collision is first executed by PYTHIA, resulting a
partonic initial state. It then goes through the partonic rescattering and the
hadronization. Eventually, the hadronic final state is obtained after hadronic
rescatterings.

For a high energy nucleus-nucleus collision in the PACIAE model~\cite{lei}, a
particle (nucleon) list is first constructed by the four- coordinate and four-momentum of all nucleons. The collision time is then calculated for each
particle pair with the straight line trajectory assumption and the requirement of
\begin{equation}
D \leq \sqrt{\sigma_{NN}^{tot} / \pi},
\label{eq1}
\end{equation}	
where the $D$ refers to the minimum approaching distance between two colliding
nucleon trajectories and $\sigma_{NN}^{tot}$ is the nucleon-nucleon (NN)
total cross section. The collision time list is then constructed.

A nucleus-nucleus collision is not described as a simple but a cumulate
superposition of the NN (a joint name for p-p, p-n, n-p, and n-n) collisions.
The simple superposition means, a nucleus-nucleus collision simulation process
ends at the empty of the collision time list. But the cumulate superposition
needs to update the particle list and the collision time list after each NN
collision. The particle list update means first removing collided nucleons
from particle list, and then adding newly generated particles into the
particle list. Similarly updating the collision time list means first removing
the collision pairs containing collision nucleons from the collision time
list, and then placing the newly formed collision pairs into the collision time list.

In the statistical mechanics~\cite{kobo} the yield of $N$ particle
(constituent) cluster (framer) reads
\begin{equation}
Y=\frac{1}{N!}\int ...\int \frac{dp_1dq_1...dp_Ndq_N}{h^{3N}}.	
\end{equation}	
For this cluster to exist naturally but not just statistically, some dynamical
constraints must be satisfied~\cite{cao}.  In order the photon pairs coalescing
to ALPs, the $\gamma\gamma \rightarrow a$ branching ratio under the $2m_K$
threshold is just the constraint.

For coalescing photon pairs in the simulated final photon list into ALPs, a
double-loop cycling over photon list (one over $i$ photon and the other over
$j$ photon, for instance) is first constructed. If the ($i-j$) photon pair
total energy is less than $2m_k$ threshold, an ALP is formed with probability of the
branching ratio of 0.57. The coalesced ALP three-momentum is the sum of three-momentum of the $i$ and $j$ constituent photons, like the coalescence
hadronization model in the AMPT~\cite{ampt} and PACIAE~\cite{lei} models.

\section{Results and Conclusions}
We have simulated five million final hadronic state (FHS) events for the
60-80 \% centrality Au+Au collisions at $\sqrt{s_{NN}}=$200 GeV by PACIAE 3.0
model~\cite{lei} with default model parameters. The calculated event averaged
ALP yield is given in the Tab.~\ref{yield}. It shows that the yield decreases
with the mass increasing whether in the full phase space or in the partial phase space of the transverse momentum ($p_T$) in [0.15, 30] GeV/c and
rapidity ($y$) in [-1.5, 1.5].

\begin{table}[htpb]
\centering	
\caption{ALP yields in the partial phase space and full phase space respectively.}
\begin{tabular}{ccccccc}
\hline
&\multicolumn{3}{c}{Partial phase space}&
\multicolumn{3}{c}{Full phase space}\\
\hline
$m_{a}$ (GeV) &0.167 &0.500 &0.833 &0.167 &0.500 &0.833  \\
Yield        &16.648 &8.615 &2.189 &28.428 &12.192 &3.237 \\
\hline
\end{tabular}
\label{yield}
\end{table}

\vspace{20pt}

\begin{figure}[htbp]
\begin{center}
\epsfig{file=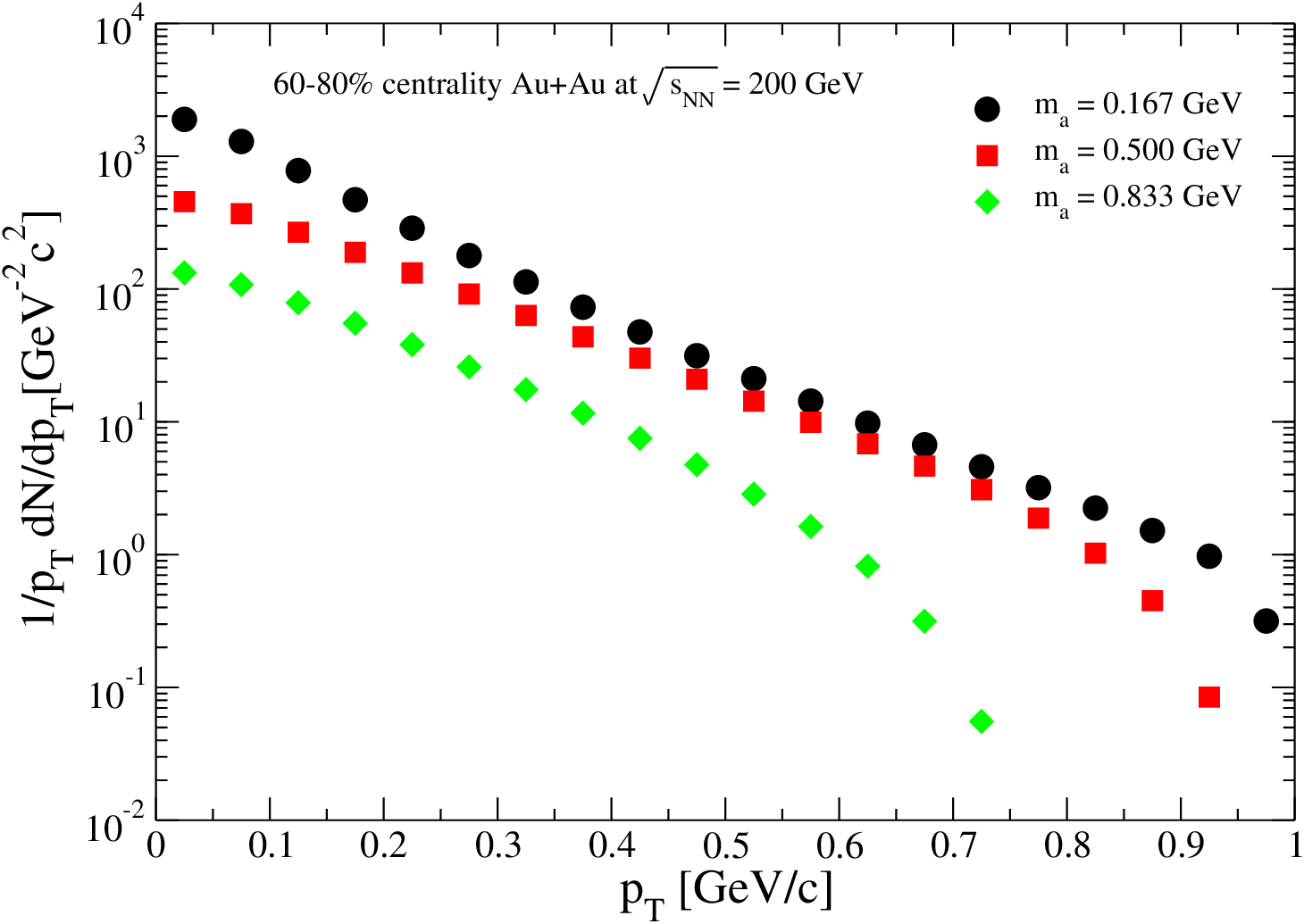,width=13.0cm,angle=0}
\caption{ALP transverse momentum distribution in the 60-80 \%
centrality Au+Au collisions at $\sqrt{s_{NN}}=$200 GeV.}
\end{center}
\label{alppt}
\end{figure}

\begin{figure}[htbp]
\begin{center}
\epsfig{file=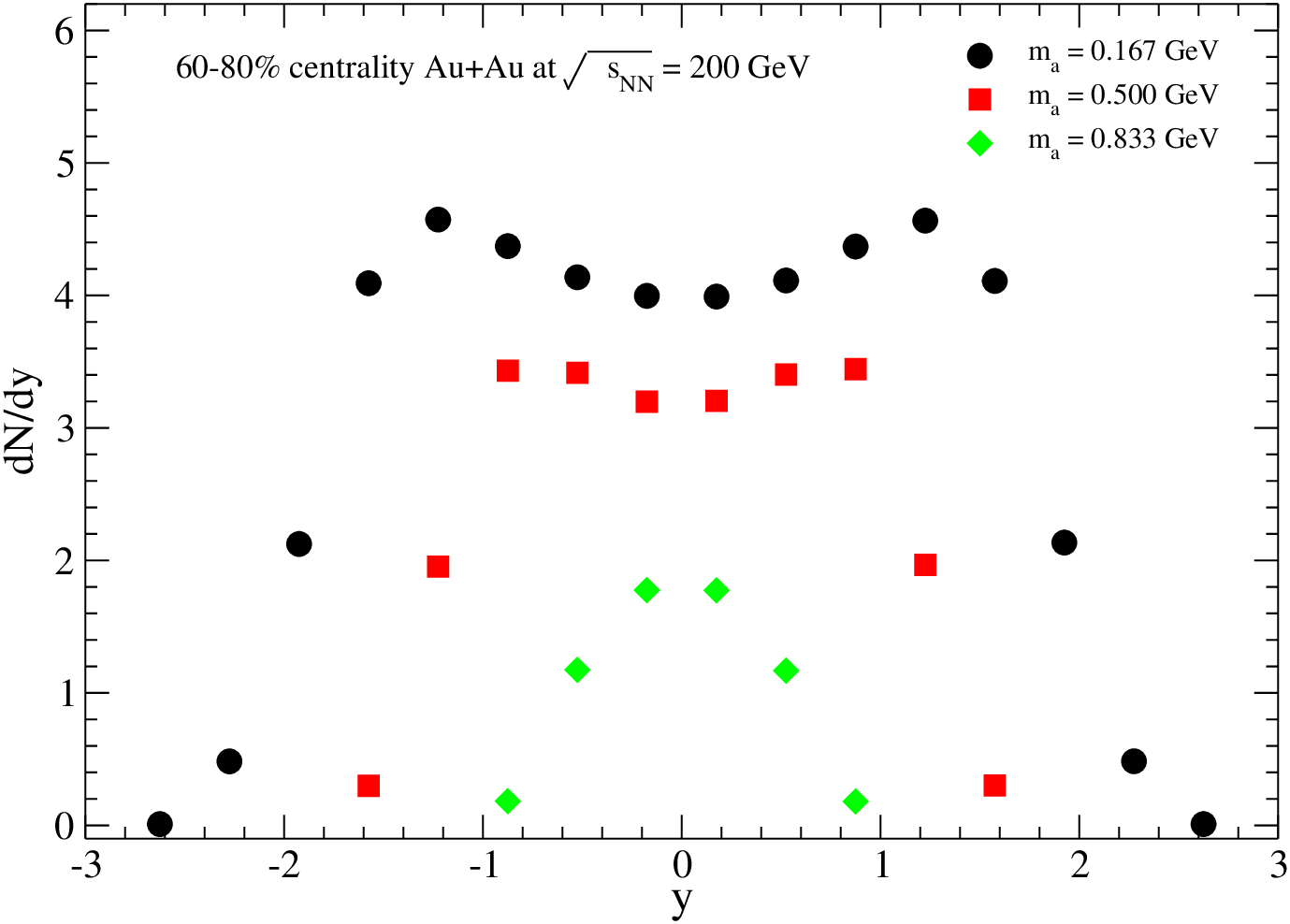,width=13.0cm,angle=0}
\caption{ALP rapidity spectra in the 60-80 \%
centrality Au+Au collision at $\sqrt{s_{NN}}=$200 GeV.}
\end{center}
\label{alpy}
\end{figure}

The Fig. (2) gives ALP transverse momentum ($p_T$) distributions.
In this figure we know that the $p_T$ distribution shape is nearly mass
independent. But the distribution magnitude decreases with the mass increasing.

We show the ALP rapidity ($y$) spectra in the Fig. (3). This figure
indicates that there are central plateaus with widths around 0.4, 2.4,
and 3.0 of the unit of rapidity for ALP mass of 0.833, 0.500, and 0.167 GeV,
respectively. The plateau width decreases with the ALP mass increasing.

In summary, we have first generated a large number of the final hadronic states
(the final photonic lists) by PACIAE model for the 60-80 \% centrality Au+Au
collisions at $\sqrt{s_{NN}}=$200 GeV. Based on these simulated final photonic
lists the ALPs are coalesced by photon pairs with total energy less than
$2m_K$ threshold and with the probability of proper branching ratio of 0.57 by
the DCPC model for the first time. We segment ALP mass into groups and count
the yield, the transverse momentum distribution and the rapidity spectrum for
each group individually. These first time theoretical predictions are worthy
to be investigated by the experiments.

In the next work we would employ the extremely rich gluons in the simulated
final partonic state to construct ALPs via $gg\rightarrow aa$. Furthermore
we shall probe the relationship between the ALP photonic- as well as
gluonic-production and the QCD phase transition, via ALP nuclear modification
factor for instance.

\section{Acknowledgments}
This work was supported by the National Natural Science
Foundation of China under Grants  No. 12375135, and by the
111 Project of the Foreign Expert Bureau of China. Y.L.Y. acknowledges the
financial support from the Key Laboratory of Quark and Lepton Physics of the Central
China Normal University under Grant No. QLPL201805 and the Continuous Basic
Scientific Research Project (Grant No. WDJC-2019-13). The work of W.C.Z. is supported
by the Natural Science Basic Research Plan in Shaanxi Province of
China(Program No. 2023-JCYB-012).

\end{document}